# Mathematically Modeling the GPe/STN Neuronal Cluster to Account for Parkinsonian Tremor and Developing a Novel Method to Accurately Diagnose Parkinson's Disease Using Speech Measurements and an Artificial Neural Network

Pooja Chandrashekar

Thomas Jefferson High School for Science and Technology
6560 Braddock Road, Alexandria VA 22312

# Abstract

Parkinson's disease (PD) is a debilitating motor system disorder characterized by progressive loss of movement, tremors, and speech slurring. PD is due to the loss of dopamine-producing brain cells, and symptoms only worsen over time, making early detection and diagnosis of the disease key to effective management and treatment. There is currently no standardized method of diagnosis available, and instead a combination of a patient's medical history and physician judgment is used. In this research, a novel method of accurately diagnosing PD using an artificial neural network (ANN) and speech measurements was developed. Using this technology, a patient need only speak into a computer microphone. Speech data is then analyzed using Praat and inputted into the ANN to obtain the diagnosis. The ANN, built using MATLAB and trained and tested with actual patient data, was able to correctly classify 96.55% of test data. A mathematical model of the GPe/STN neural cluster was then constructed to account for Parkinsonian tremor. This was done using a two-cell model that coupled a GPe neuron to a STN neuron and consisted of a system of twelve ordinary differential equations. The model was structured to account for neuron bursting behavior, neurotransmitter release, and synaptic connections between neurons. After comparison to the biological behavior of the cluster and neuron firing patterns in PD patients, the model was determined to be predictive of known biological results. This breakthrough work presents a significant step forward in PD research and could be successfully implemented into clinical practice.

# 1  Introduction

Parkinson's disease (PD) is a movement disorder that stems from the low concentration of dopamine in Parkinson's affected neurons in the basal ganglia and the consequent dysfunction of basal ganglia circuitry, resulting in neurological conditions such as PD. In this work, we attempted to both address the problem of a current lack of a standardized method of diagnosis of PD as well as understand the mechanism by which Parkinsonian tremor may arise.

  Parkinsonian tremor, one of the defining characteristics of PD, is marked by significant speech slurring and atypical speech measurements. It was thus hypothesized that the GPe/STN neuronal cluster could be mathematically modeled to account for Parkinsonian tremor since previous scientific work has shown that the oscillatory nature of this basal ganglia cluster may be responsible for Parkinsonian tremor.Prior work has shown that the cluster is comprised of "leaky integrate and fire" neurons which exponentially decay due to PD neurodegeneration while also being subjected to external excitatory or inhibitory impulses. It was hypothesized that a system of ordinary differential equations could be developed to produce both irregular and correlated activitywithin the network, as well as satisfy known computational and biological results regarding various ion currents.

  In order to develop a simple and accurate method of diagnosis for PD, a novel method of using an artificial neural to accurately diagnose PD using speech measurements collected and analyzed via a microphone was researched. There is currently no available laboratory or automated method of diagnosing PD apart from a careful examination of a patient's medical history, often making it difficult to accurately diagnose and detect PD in its earlier stages. It was hypothesized that a neural network can be built to take input from an attached microphone, analyze and determine specific speech measurements using phonetic analysis software, and then output whether or not the patient has PD based upon results obtained from comparing the speech measurements with those used to train the network. Successful implementation of this network into clinical practice would be a key step forward in developing a simple and effective method of PD diagnosis.

## 1.1 Parkinson's Disease

Parkinson's disease is classified as a progressive movement disorder that originates in the basal ganglia and is brought about as a result of the decreased stimulation of the motor cortex by the basal ganglia. This decreased stimulation results from the lack of formation of dopamine by the substantianigra pars compacta (SNc). This lack of stimulation is also responsible for the consequent inhibition of the thalamus, which leads to muscle rigidity, Parkinsonian tremor, akinesia (difficulty starting or stopping movement), bradykinesia (slowing of movement), and postural instability.

      The GPe/STN neural cluster, a network of GPe and STN-type neurons locate in the basal ganglia, has been found to behave differently during normal and Parkinsonian states. In normal conditions, neuron firing is largely uncorrelated and displays no significant spatiotemporal pattern. However, in Parkinsonian states, neuron firing is shown to be highly correlated and synchronous, although the reason for this synchronous bursting is unknown. Prior scientific work has also suggested that oscillations in this particular neuronal cluster may be responsible for Parkinsonian tremor, although the mechanism by which oscillations arise and contribute to Parkinsonian tremor is not yet understood.

      One of the key characteristics of PD is the lack of a standardized system of diagnosis. Currently, not tests exist to diagnose Parkinson's disease, severely impeding the development of a fluid patient-tailored approach to PD treatment and therapy that relies upon early diagnosis of the disease. In order to diagnose PD, physicians now use a combination of a patient's medical history, review of symptoms, and a neurological and physical examination. The diagnosis process is often time consuming and highly inconvenient, often requiring several clinical visits as well as various tests or medications.

## 1.2 Behavior of the GPe/STN Neural Cluster

The basal ganglia are a group of five nuclei that communicate with the thalamus, brainstem, and cerebral cortex to regulate motor control, cognition, and learning. The GPe/STN network is a group of neurons located in the basal ganglia that is involved in the indirect pathway of the basal ganglia. The GPe and STN neurons, the two major classes of neurons that comprise the network, form a feedback system termed the Central Pacemaker Generator (CPG) due to the high level of connectivity between the nuclei of these neurons. The CPG can remain non-quiescent without any external input from their neural regions, meaning that striatal input only serves to modulate or regulate the firing patterns of the cluster.

The GPe/STN neural cluster has shown to behave differently during normal and Parkinsonian states. During the normal state, neuron firing behavior has been found to be uncorrelated and displays no significant spatiotemporal pattern. However, during the Parkinsonian state, neuron firing behavior has found to be highly correlated and synchronous, although the reason for this synchronous bursting behavior is still unknown. Figure 1 shows the basal ganglia circuitry and dysfunction in this circuitry can result in neurological conditions such as PD.

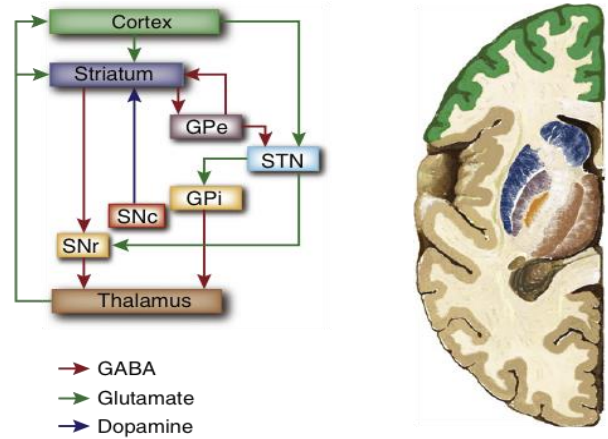

Figure 1. Basal ganglia circuitry.

The mathematical model was constructed so as to agree with known computational and biological results from existing literature. While similar types of currents are associated with both the STN and GPe neurons, some of these currents differ in their behavior and magnitude between the two. For both neurons, an ohmic leak current $I_L$ is present due to open resting channels through which current can escape the capacitor system. The STN neuron has two types of potassium currents, collectively represented $I_{Ks}$, both a persistent outward potassium current that is activated by depolarization and which is responsible for limiting the duration of an action potential, as well as a transient outward current that is activated quickly at low thresholds and which causes more complex and uncorrelated firing patterns such as periodic bursting. Similarly, the GPe neuron also has low and high threshold potassium currents $I_{Kg}$. The two neurons differ in their afterhyperpolarization currents as the afterhyperpolarization current $I_{AHPs}$ of the STN neuron is a calcium-dependent potassium current that may be responsible for the slow afterhyperpolarization characteristic of STN neurons, while the $I_{AHPg}$ of the GPe neuron may be responsible for both the fast and slow afterhyperpolarization of GPe neurons. In addition, while the sodium current $I_{Nas}$ for the STN neuron is a transient inward current that allows for the generation of action potentials and spontaneous firing, the sodium current $I_{Nag}$ for the GPe neuron is a persistent inward current. Finally, both the calcium currents $I_{Cas}$ for the STN neuron and the calcium currents $I_{Cag}$ for the GPe neuron consist of various inward currents responsible for the plateau potential and the low-threshold $I_T$ current that produces the rebound potential. The model was structured to account for these various types of currents and provide a simplified framework for understanding how they interact within the system, particularly focusing on neurotransmitter binding, the gating mechanisms of receptor channels, and the change in internal calcium concentrations of the neurons. All of these currents, along with the currents between the two neurons $I_{g\ to\ s}$ and $I_{s\ to\ g}$ due to the nature of their synaptic coupling, were accounted for in the capacitance of the membrane and these particular characteristic were chosen because neurotransmitter behavior and calcium concentrations play a key role in regulating the behavior of the GPe/STN neural network. In addition, the kinetic model of neurotransmitter binding $A + B \leftrightarrow BA$ was used.

## 2 Methods and Techniques

There were two phases to this research, with Phase 1 being the mathematical model of the GPe/STN neural cluster and Phase 2 being the artificial neural network for the diagnosis of PD. Each of these phases is described separately in the following two subsections.

### 2.1 Two-Cell Mathematical Model

The goal of Phase 1 was to mathematically model the GPe/STN neuronal cluster located in the basal ganglia to determine whether interactions within this system can account for Parkinsonian tremor by producing both correlated and irregular behavior for varying parameters over time using a set of differential equations. This model was largely theoretical in that no experimental data regarding the behavior of the cluster is currently available, so parameter and constant values must be determined after experimentation.This computation model was structured using the NEURON simulation environment and the mathematical modeling was completed using the MATLAB Differential Equation Toolbox.

Existing computational models of the basal ganglia and striatal were first simplified to a simple two-cell model consisting of only a single GPe neuron and a single STN neuron, as shown in Figure 2. External inputs, such as those from the striatum and thalamus, were not taken into account in this model.

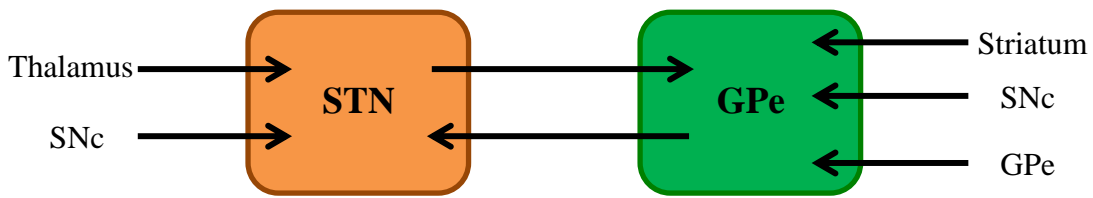

Figure 2.Two-cell model of the GPe/STN neural cluster.

For the purposes of this model, the cell membrane was treated as a capacitor responsible for separating the charge between the inside and outside biological mediums, as shown in Figure 3. After the appropriate current parameters around which to build the model around were selected, these currents were incorporated into the model by using resistors in parallel to represent the membrane conductance parameters $g_K$, $g_{Na}$, and $g_L$. These "resistors" would then effectively create currents in parallel that would produce individual capacitances which additively sum to the overall capacitance $C_m$ of the membrane. Thus, we have $I_{total} = V + \sum I$.

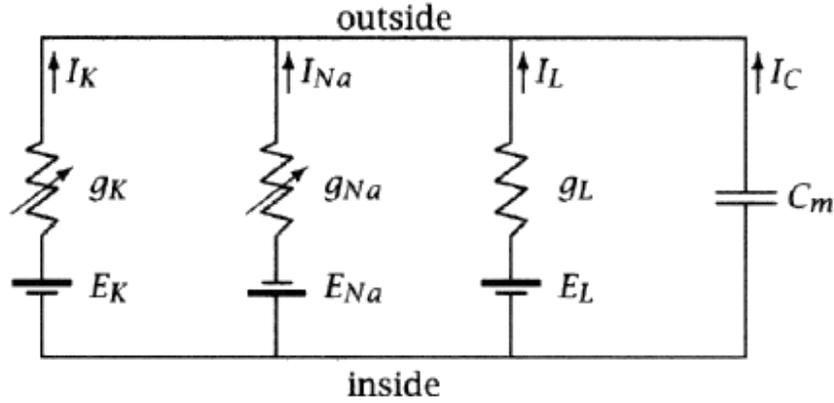

Figure 3. Treating the cell membrane as a capacitor.

Models for the STN and GPe neurons were then built separately and each was structured around the well-known Hodgkin-Huxley conductance-based model that describes how action potentials are propagated throughout a series of neurons. This was done by considering the sodium, potassium, calcium, afterhyperpolarization, and synaptic currents as contributing to the charge separation of the membrane.

The two-cell model of the GPe/STN neural cluster was constructed by accounting for the synaptic coupling on the two neurons in the network. This was done by using ordinary differential equations (ODEs) to model the relationship between the action potentials in the presynaptic cell and the gating of the receptor channels in the postsynaptic cell. The synaptic currents between the neurons and the fraction of open channels were set as variables. To construct the final model, both ODEs for the gating variables of the postsynaptic receptor channels and the internal calcium concentration were considered, in addition to the synaptic coupling ODEs built previously. Finally, analogous ODEs for both the GPe and the STN neurons were constructed, as shown below.

The following ODEs represent treating the membrane as a capacitor with the inner and outer membrane as the two conductors responsible for the charge separation. The capacitative current is the negative of the sum of the separate currents, including currents of potassium, sodium, and calcium ions, an afterhyperpolarization current, a current provided by neurotransmitter release, an ohmic leak current, and a current between the two neurons.

$$C_m \frac{dV_s}{dt} = -I_{Ks} - I_{Nas} - I_{Cas} - I_{AHPs} - I_{Ts} - I_{Ls} - I_{g\,to\,s}$$

$$C_m \frac{dV_g}{dt} = -I_{Kg} - I_{Nag} - I_{Cag} - I_{AHPg} - I_{Tg} - I_{Lg} - I_{s\,to\,g}$$

The following ODEs represent the gating variables $n$, $h$, and $r$ for both the GPe and STN neurons. In all of these equations, the $A_\infty(V)$ is the steady-state function when neurotransmitter binding stabilizes over time, and the ODE simply tells us the change in the gating of the receptor channels over time where $\tau_A$ is the time over which the neurotransmitter binding occurs.

$$\frac{dn_s}{dt} = \emptyset n_s \frac{n_{\infty s}(V_s) - n_s}{\tau_{ns}(V_s)} \qquad \frac{dn_g}{dt} = \emptyset n_g \frac{n_{\infty g}(V_g) - n_g}{\tau_{ng}(V_g)}$$

$$\frac{dh_s}{dt} = \emptyset h_s \frac{h_{\infty s}(V_s) - h_s}{\tau_{hs}(V_s)} \qquad \frac{dh_g}{dt} = \emptyset h_g \frac{h_{\infty g}(V_g) - h_g}{\tau_{hg}(V_g)}$$

$$\frac{d\tau_s}{dt} = \emptyset \tau_s \frac{r_{\infty s}(V_s) - r_s}{\tau_{\tau s}(V_s)} \qquad \frac{d\tau_g}{dt} = \emptyset \tau_g \frac{r_{\infty g}(V_g) - r_g}{\tau_{\tau g}(V_g)}$$

The following ODEs represent the change in the internal calcium concentration of the two neurons over time.

$$\frac{d[Ca]_s}{dt} = \epsilon_s(-I_{Cas} - I_{Ts} - kCa_s[Ca]_s) \qquad \frac{d[Ca]_g}{dt} = \epsilon_g(-I_{Cag} - I_{Tg} - kCa_g[Ca]_g)$$

The following ODEs represent the synaptic coupling of the two neurons, particularly the change in time of the fraction of open receptor channels. $H$ is the Heaviside function.

$$\frac{df_{s\,to\,g}}{dt} = \alpha_g H(v_s - \theta_g)(1 - f_{s\,to\,g}) - \beta_{s\,to\,g}$$

$$\frac{df_{g\,to\,s}}{dt} = \alpha_s H(v_g - \theta_s)(1 - f_{g\,to\,s}) - \beta_{g\,to\,s}$$

## 2.2 Feedforward Artificial Neural Network

The goal of Phase 2 was to build an artificial neural network capable of diagnosing PD based on spoken speech and subsequent speech analysis as there are currently no available tests to diagnose PD other than a careful examination of a patient's medical history. The data used to build the neural network was obtained from the UC Irvine Data Repository and the dataset used consisted of 197 biomedical voice measurements from both patients diagnosed with PD and healthy individuals. An excerpt of the dataset is shown in Figure 4. The parameters that were used to build the network included the average vocal fundamental frequency, maximum vocal fundamental frequency, minimum vocal fundamental frequency, jitter (variation in fundamental frequency), shimmer (variation in amplitude), noise-to-harmonics ratio, harmonics-to-noise ratio, and the health status of the patient. The goal of the artificial neural network (ANN) was to discriminate between patients with PD and healthy individuals. The feedforward artificial neural network was built and coded using MATLAB and the obtained data was used to train, validate, and test the neural network. The dataset was divided as follows for the three basic stages of machine learning: 70% for training, 15% for validation, and 15% for testing. The backpropagation algorithm, a common tenet of machine learning and pattern recognition techniques, was used to program the feedforward artificial neural network.

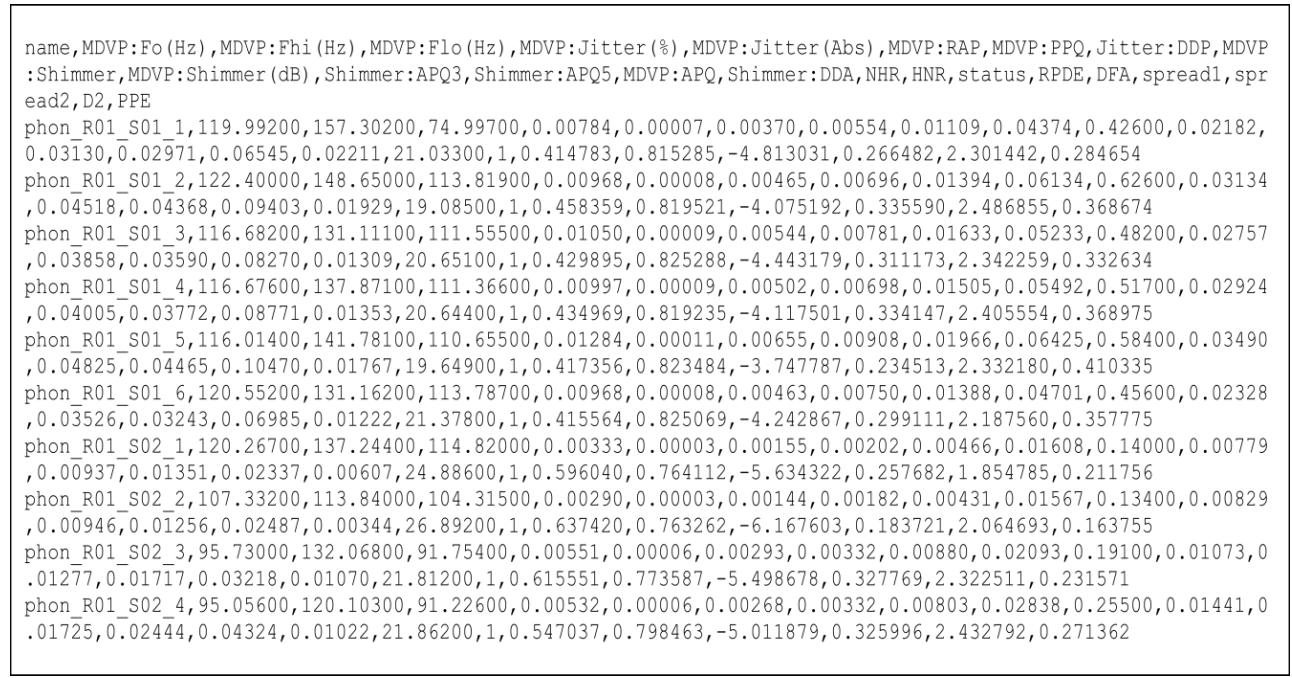

```
name,MDVP:Fo(Hz),MDVP:Fhi(Hz),MDVP:Flo(Hz),MDVP:Jitter(%),MDVP:Jitter(Abs),MDVP:RAP,MDVP:PPQ,Jitter:DDP,MDVP
:Shimmer,MDVP:Shimmer(dB),Shimmer:APQ3,Shimmer:APQ5,MDVP:APQ,Shimmer:DDA,NHR,HNR,status,RPDE,DFA,spread1,spr
ead2,D2,PPE
phon_R01_S01_1,119.99200,157.30200,74.99700,0.00784,0.00007,0.00370,0.00554,0.01109,0.04374,0.42600,0.02182,
0.03130,0.02971,0.06545,0.02211,21.03300,1,0.414783,0.815285,-4.813031,0.266482,2.301442,0.284654
phon_R01_S01_2,122.40000,148.65000,113.81900,0.00968,0.00008,0.00465,0.00696,0.01394,0.06134,0.62600,0.03134
,0.04518,0.04368,0.09403,0.01929,19.08500,1,0.458359,0.819521,-4.075192,0.335590,2.486855,0.368674
phon_R01_S01_3,116.68200,131.11100,111.55500,0.01050,0.00009,0.00544,0.00781,0.01633,0.05233,0.48200,0.02757
,0.03858,0.03590,0.08270,0.01309,20.65100,1,0.429895,0.825288,-4.443179,0.311173,2.342259,0.332634
phon_R01_S01_4,116.67600,137.87100,111.36600,0.00997,0.00009,0.00502,0.00698,0.01505,0.05492,0.51700,0.02924
,0.04005,0.03772,0.08771,0.01353,20.64400,1,0.434969,0.819235,-4.117501,0.334147,2.405554,0.368975
phon_R01_S01_5,116.01400,141.78100,110.65500,0.01284,0.00011,0.00655,0.00908,0.01966,0.06425,0.58400,0.03490
,0.04825,0.04465,0.10470,0.01767,19.64900,1,0.417356,0.823484,-3.747787,0.234513,2.332180,0.410335
phon_R01_S01_6,120.55200,131.16200,113.78700,0.00968,0.00008,0.00463,0.00750,0.01388,0.04701,0.45600,0.02328
,0.03526,0.03243,0.06985,0.01222,21.37800,1,0.415564,0.825069,-4.242867,0.299111,2.187560,0.357775
phon_R01_S02_1,120.26700,137.24400,114.82000,0.00333,0.00003,0.00155,0.00202,0.00466,0.01608,0.14000,0.00779
,0.00937,0.01351,0.02337,0.00607,24.88600,1,0.596040,0.764112,-5.634322,0.257682,1.854785,0.211756
phon_R01_S02_2,107.33200,113.84000,104.31500,0.00290,0.00003,0.00144,0.00182,0.00431,0.01567,0.13400,0.00829
,0.00946,0.01256,0.02487,0.00344,26.89200,1,0.637420,0.763262,-6.167603,0.183721,2.064693,0.163755
phon_R01_S02_3,95.73000,132.06800,91.75400,0.00551,0.00006,0.00293,0.00332,0.00880,0.02093,0.19100,0.01073,0
.01277,0.01717,0.03218,0.01070,21.81200,1,0.615551,0.773587,-5.498678,0.327769,2.322511,0.231571
phon_R01_S02_4,95.05600,120.10300,91.22600,0.00532,0.00006,0.00268,0.00332,0.00803,0.02838,0.25500,0.01441,0
.01725,0.02444,0.04324,0.01022,21.86200,1,0.547037,0.798463,-5.011879,0.325996,2.432792,0.271362
```

Figure 4. Excerpt of speech data for 10 patients.

In order to obtain the appropriate voice measurements needed for the neural network to be able to diagnose PD based on spoken speech using an attached microphone, the speech waveforms must first be analyzed. This was done using Pratt acoustic analysis software in conjunction with the MATLAB speech analysis toolbox. A Praat script was written to take input from a computer microphone and analyze the waveform so as to determine the appropriate speech measurements needed for comparison against the dataset used to train the network. The speech analysis program was then linked to the artificial neural network to allow the MATLAB program to process microphone input and display patient diagnosis and probability of accuracy. The probability of accuracy displayed by the MATLAB program was not the accuracy of the network itself, but rather the probability that a particular diagnosis for a particular speech recording is correct. The speech recordings used to test the network were restricted to those in which the patient spoke a long vowel sound since this type of recording has been shown to produce the best results when analyzing spoken speech for PD in previous work. Finally, error histograms, confusion matrices, performance plots, training state plots, and receiver operating characteristic plots were created to display the error probabilities of the constructed neural network. The complete information flow between the various programs and entities in Phase 2 is shown in Figure 5, with the next step towards public release being the development of a web applet to make the entire process more fluid and accessible.

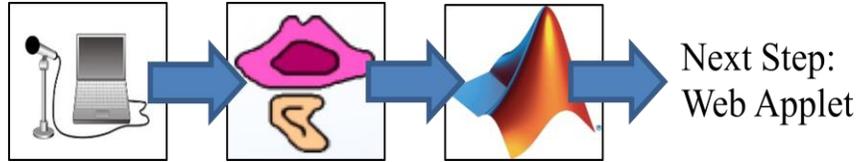

Figure 5. Information flow from microphone to Praat to MATLAB to web applet.

## 3 Results and Analysis

The mathematical model was designed based upon the Hodgkin-Huxley model and was built so as to account for known computational and biological results regarding the various currents considered in this conductance-based model in which the cell membrane was treated as a capacitor. This model was largely theoretical due to the fact that no experimental results of the GPe/STN neural network currently exist against which the model could be compared and analyzed. A system of 12 ODEs was developed, focusing on neurotransmitter binding based on the fraction of open receptor channels and gating variables, internal calcium concentration, and synaptic coupling of the two neurons.

The pattern recognition artificial neural network built was found to accurately classify 96.55% and inaccurately classify 3.45% of speech data, which consisted of both healthy and PD patient measurements. The neural network was programmed using a feedforward backpropagation neural network algorithm and implemented in MATLAB. Figure 6 shows the output of a sample Praat speech analysis script run on a voice recording of a patient with PD, along with the MATLAB output for this speech sample which shows that the program was able to accurately diagnose this patient with PD and also provided a probability of accuracy regarding the probability that this particular diagnosis was accurate. The overall accuracy of the network of 96.55% is a significant improvement over the few other attempts to develop an artificial neural network to diagnose PD which only reached an accuracy of approximately 80%.

The neural network was found to give best results with 7 hidden neurons, so the ANN was programmed with 7

```
Time range of SELECTION
   From 2.160342 to 4.109046 seconds
(duration: 1.948704 seconds)
Pitch:
   Median pitch: 259.216 Hz
   Mean pitch: 257.829 Hz
   Standard deviation: 16.663 Hz
   Minimum pitch: 218.940 Hz
   Maximum pitch: 294.502 Hz
Jitter:
   Jitter (local): 0.399%
   Jitter (local, absolute): 15.496E-6 seconds
   Jitter (rap): 0.123%
   Jitter (ppq5): 0.158%
   Jitter (ddp): 0.370%
Shimmer:
   Shimmer (local): 9.063%
   Shimmer (local, dB): 0.866 dB
   Shimmer (apq3): 3.901%
   Shimmer (apq5): 5.800%
   Shimmer (apq11): 9.203%
   Shimmer (dda): 11.702%
Harmonicity of the voiced parts only:
   Mean autocorrelation: 0.944773
   Mean noise-to-harmonics ratio: 0.061250
   Mean harmonics-to-noise ratio: 13.661 dB
```

```
Diagnosis: Patient does have PD
Probability of Accuracy: 69.33%
```

Figure 6. Praat speech analysis output (top) and MATLAB output for speech sample (bottom).

hidden neuron layers so as to provide optimal predictive validity. This is shown in the network schematic diagram in Figure 7.

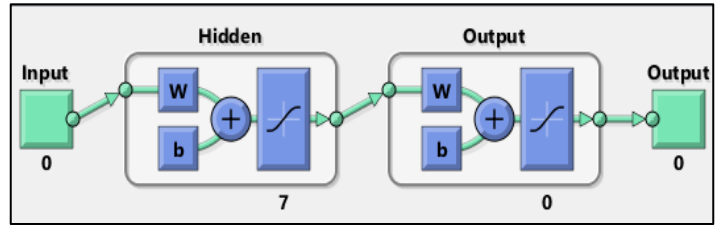

Figure 7. Network schematic diagram.

The best validation performance plot in Figure 8 shows that the mean square error (MSE) does slowly decrease over the course of the network training state. Although there was not a large decrease in the training MSE, the results of the performance plot do show that the validation and test data produced very similar results to the training dataset, thus validating the performance of the network.

The total confusion matrix in Figure 9 shows the percentage of false positives (9.7%), true positives (75.4%), false negatives (0.0%), and true negatives (14.9%).

The ROC plot in Figure 11 also shows a generally low error with the receiver operating characteristic curve far from the line of random guessing, although the plot does show that some error does arise during the course of testing and validation of the data.

The error histogram in Figure 10 shows that the error data clustered around zero error, although the histogram is skewed to the left due to the result of a high number of false positives compared to the extremely low number of false negatives. In addition, the majority of the test data is located around zero error, with only a few test values falling to the left and right.

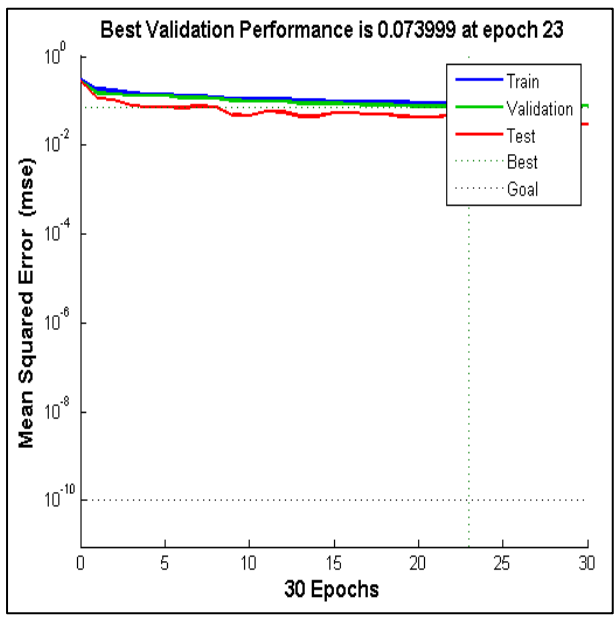
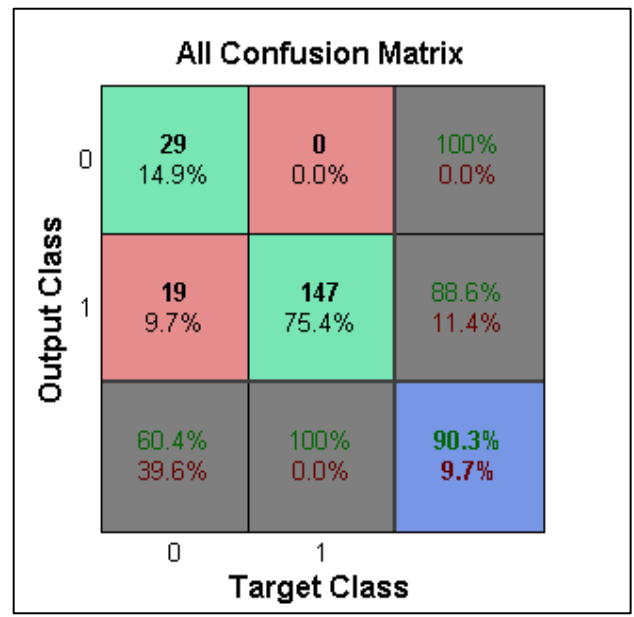

Figure 8. Best validation performance plot.    Figure 9. Total confusion matrix.

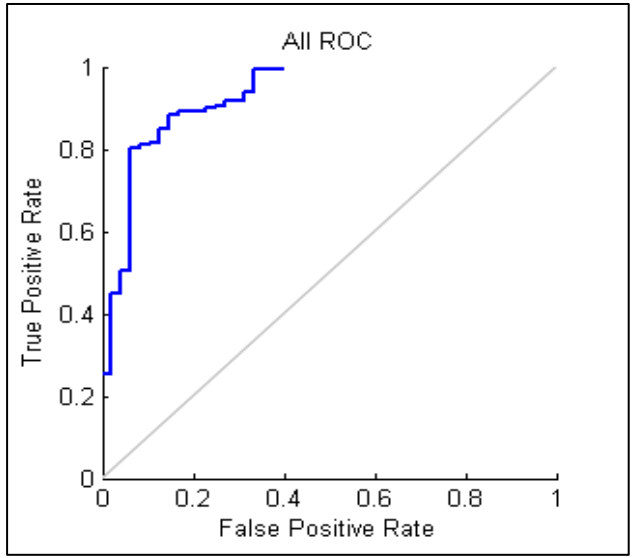 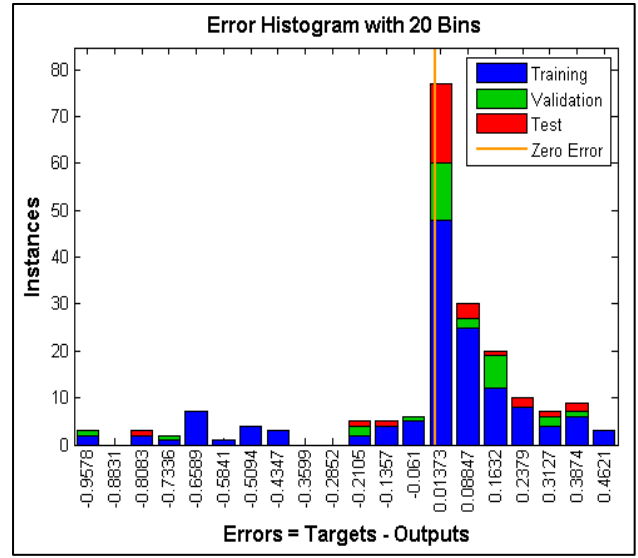

Figure 10. ROC plot.  Figure 11. Error histogram.

## 4 Conclusions and Implications

The novel mathematical model developed in this research shows that oscillations in the GPe/STN neural cluster are responsible for Parkinsonian tremor, one of the defining characteristics of PD. This model offers new insights into the mechanism by which these tremors may arise by presenting these oscillations as a combination of different parameters such as sodium and potassium concentrations, calcium ion concentrations, a current provided by neurotransmitter release, an ohmic leak current, and a current between the two neurons. This model is especially significant because it provides a much more simplified model of the basal ganglia neural cluster as compared to existing models of the basal ganglia that are highly complicated and require many more ODEs to account for the high number of variables. Although the number of neurons in this model must be further extended to gain a complete understanding of the manner by which oscillations in the cluster arise, the simple framework of this conductance-based model based upon treating the cell membrane as a capacitor responsible for charge separation will prove invaluable in constructing a larger model.

     One of the most promising results of this research is the novel method of diagnosis using an artificial neural network and speech measurements that was developed. Very little research has been done in the area of using speech measurements to diagnose PD and the few international efforts that have been made have proved largely unsuccessful due to low accuracy of results. This work presents the first highly accurate diagnosis system that requires the patient to only speak a long vowel sound into an attached microphone for the program to output the diagnosis. Because it is especially important that PD be diagnosed early due to its progressive nature, this technology would allow for patients to be diagnosed more accurately and earlier so that patient-tailored therapy and treatment could be administered optimally. In addition, while current diagnosis often requires multiple clinical visits and medications, this system can

diagnosis PD quickly, is accurate, and can be released to the public at little to no cost. This research thus presents a significant step forward in the diagnosis of PD and the technology developed has the potential to be implemented into clinical practice, as well as released as a web application that could be made freely available to the public for home use.

Further research needs to be conducted in order to establish experimental results for the behavior of the GPe/STN neuronal cluster. Investigating the behavior of the cluster and obtaining such data would greatly improve the accuracy of the model and allow for it to be modified as needed to obtain the best possible approximation for experimental results. This approximation could then be analyzed and plotted to provide visual representations of the accuracy of the model so as to better understand the effect of varying model parameters. Experimental data would be necessary in order to establish model parameters and constant values. The model should also be expanded to include a larger network of GPe and STN neurons and the different types of GPe and STN neurons should be treated differently so as to account for small variations in their behavior. In addition, striatal input and the current between two neurons of the same type should be taken into consideration.

One of the key steps that must be taken to increase the predictive accuracy of the neural network is to test the technology with actual PD patients and healthy volunteers. This would help test whether the process of collecting speech measurements using Praat and inputting this information into the neural network needs to be facilitated or improved. It would also provide more training data to improve the accuracy of the network. The reasonable next step to furthering the neural network technology is to develop a web or mobile application for global public usage. The only additional equipment individuals would need is a microphone which could be used to create a sound file to load into the application, which would then provide an accurate diagnosis within seconds. The increased usage of the network upon releasing the web application to the public would help train the artificial neural network as well, so that over time, the accuracy of the neural network would continue to increase as it learned the subtleties of diagnosis.

# 5  References

Bishop, Christopher M. *Neural Networks and Machine Learning*. Berlin: Springer, 1998. Print.
Cutsuridis, V., and S. Perantonis. "A Neural Network Model of Parkinson's Disease
    Bradykinesia." *Neural Networks* 19.4 (2006): 354-74. Print.
Dayan, Peter, and L. F. Abbott. *Theoretical Neuroscience: Computational and
    Mathematical Modeling of Neural Systems*. Cambridge, MA: Massachusetts
    Institute of Technology, 2001. Print.
"Diagnosis of Parkinson's Disease." University of Maryland Medical Center, 24 May
    2013. Web. 17 Dec. 2013.
'Exploiting Nonlinear Recurrence and Fractal Scaling Properties for Voice Disorder Detection',
    Little MA, McSharry PE, Roberts SJ, Costello DAE, Moroz IM. BioMedical Engineering
    OnLine 2007, 6:23 (26 June 2007).
Gil, David A., and Magnus B. Johnson. "Diagnosing Parkinson by Using Artificial

Neural Networks and Support Vector Machines." *Global Journal of Computer Science and Technology* (n.d.): 63-71. Print.

Khemphila, Anchana, and Veera Boonjing. "Parkinsons Disease Classification Using Neural Network and Feature Selection." *World Academy of Science, Engineering and Technology* (2012): n. pag. Print.

Rich, Scott, and Michael Reed. *Mathematical Model of the Basal Ganglia and Oscillatory Neuron Clusters in Relation to Parkinsonian Tremors*. Tech. Duke University, n.d. Web. 17 Dec. 2013.

"Tests and Diagnosis for Parkinson's Disease." *Mayo Clinic*. Mayo Foundation for Medical Education and Research, 11 May 2012. Web. 15 Dec. 2013.

Trappenberg, Thomas P. *Fundamentals of Computational Neuroscience*. Oxford: Oxford UP, 2002. Print.